\documentclass[aps,prd,floatfix,preprint,nofootinbib]{revtex4-1}
\usepackage{graphicx}
\usepackage{epic}
\usepackage{eepic}
\usepackage{latexsym}

\newcommand{\eq}[1]{(\ref{#1})}
\newcommand{\be}{\begin{equation}}
\newcommand{\ee}{\end{equation}}
\newcommand{\bea}{\begin{eqnarray}}
\newcommand{\eea}{\end{eqnarray}}

\newcommand{\hs}[1]{\hspace{#1 mm}}

\def\a{\alpha}

\def\cc{\gamma}

\def\d{\delta}

\def\e{\epsilon}

\def\fr{\frac}

\def\vf{\varphi}

\def\m{\mu}
\def\n{\nu}

\def\r{\rho}
\def\s{\sigma}

\def\O{\Omega}

\def\o{\omega}

\def\vf{\varphi}

\def\del{\partial}

\let\bm=\bibitem
\def\nn{\nonumber}

\begin{document}

\title{Cosmological backreaction of a quantized massless scalar field}   

\author{Ali Kaya}
\email[]{ali.kaya@boun.edu.tr}
\author{Merve Tarman}
\email[]{merve.tarman@boun.edu.tr}
\affiliation{Bo\~{g}azi\c{c}i University, Department of Physics, \\ 34342,
Bebek, \.Istanbul, Turkey}

\date{\today}

\begin{abstract}

We consider the backreaction problem of a quantized minimally coupled massless scalar field in cosmology. The adiabatically regularized stress-energy tensor in a general  Friedmann-Robertson-Walker background is approximately evaluated by using the fact that  subhorizon modes evolve adiabatically and  superhorizon modes are frozen. The vacuum energy density is verified to obey a new first order differential equation depending on a dimensionless parameter of order unity, which calibrates subhorizon/superhorizon division. We check the validity of the approximation by calculating the corresponding vacuum energy densities in fixed backgrounds, which are shown to agree with the known results in de Sitter space and space-times undergoing power law expansions. We then apply our findings to slow-roll inflationary models. Although backreaction effects are found to be negligible during the near exponential expansion, the vacuum energy density generated during this period might be important at later stages since it decreases slower than radiation or dust. 

\end{abstract}

\maketitle

\section{Introduction}

Quantum field theory in curved space is a well established subject of study (see e.g. \cite{b1,b2,b3}). For free fields in a general background, difficulties related to the absence of Poincare symmetry, non-uniqueness of vacuum and lack of particle states were all solved and a physically viable picture has been obtained. Suitable regularization techniques, like adiabatic \cite{ad1} (see also \cite{ad2}) and point-splitting \cite{ps},  were developed to obtain finite and physically relevant expressions for the stress-energy-momentum tensor vacuum expectation values. Adiabatic and point-splitting regularizations were shown to be equivalent  in cosmological spacetimes \cite{eq1,eq2}. 

Although quantization in a {\it fixed} background is well understood, the backreaction of a quantum field on the geometry is not fully determined. For example, the cosmic evolution of the scale factor of a universe dominated by the stress-energy tensor of a quantum field is not known. The backreaction problem is difficult to tackle (see e.g. \cite{br0,br1,br11,br2,br3,br4} for some attempts relevant for cosmology) and it is expected to shed further light on semiclassical quantum gravity.  

In this paper, we try to determine the cosmological backreaction  of a minimally coupled  quantized real massless scalar field. As we will see, after adiabatic regularization the problem reduces to an integro-differential equation system involving the scale factor of the universe $a(t)$ and the mode function of the field $\m_k(t)$. In an earlier work \cite{ali}, we simplified this equation system by  using adiabatic subtraction terms up to second order and approximately evaluating the momentum integrals from the behavior of the subhorizon and the superhorizon mode functions. In that way we managed to evaluate the mode function momentum integrals in terms of the vacuum energy density and obtained analytical solutions. 

We would like to improve the method employed in \cite{ali} in various ways to have a better approximation.  First, we utilize the full adiabatic regularization, i.e. we include adiabatic order four subtraction terms to remove all the divergences in the quantum stress-energy tensor. Second, we introduce a dimensionless parameter $\a$ of order unity characterizing subhrorizon/superhorizon division of modes. Finally, we also keep finite adiabatic subtraction terms, which vanish when one naively takes the massless limit before evaluating the momentum integrals (for example these terms must be kept to obtain the trace anomaly using adiabatic regularization \cite{tr}). As a result, the final equation system involving the scale factor and the vacuum energy density  turns out to be  more complicated than the one found in \cite{ali} and it is difficult to obtain analytical solutions. However, one can still obtain quantitative information on how vacuum energy density affects the cosmic evolution. 

To check the validity of our approximation we determine the vacuum energy density in a fixed background to compare our findings with the known results in de Sitter space and space-times undergoing power law expansions. We find complete agreement in both cases by adjusting the above mentioned dimensionless parameter $\a$.  We also apply our formulas to slow-roll inflationary models and try to pin down possible backreaction effects. We show that  the vacuum energy density becomes negligible during the near exponential expansion, but  it may have a cosmological  imprint at later stages since it decreases more slower than radiation or dust.

\section{Quantization and regularization}

We consider a free massless real scalar field $\phi$ obeying 
\be
\nabla^2\phi=0.
\ee
The scalar is propagating in a Friedmann-Robertson-Walker (FRW) space-time which has the metric
\bea
ds^2&=&-dt^2+a(t)^2(dx^2+dy^2+dz^2)\nn\\
&=&a(\eta)^2(-d\eta^2+dx^2+dy^2+dz^2),
\eea
where $t$ and $\eta$ denote the proper and conformal cosmic times, respectively. We denote the corresponding Hubble parameters as 
\be 
h=\fr{a'}{a},\hs{10}H=\fr{\dot{a}}{a},
\ee
where the prime and the dot represent derivatives with respect to $\eta$ and $t$, respectively. Defining a new field $\m$ by
\be
\m=a \,\phi
\ee
and using the standard canonical quantization procedure with the action $S=\int \sqrt{-g}\, (\nabla\phi)^2$, one can decompose the field operator as 
\be
\m=\int \fr{d^3
k}{(2\pi)^{3/2}}\left[\m_k(\eta)\,e^{i\vec{k}.\vec{x}}\,a_{\vec{k}}
+\m_k(\eta)^*\,e^{-i\vec{k}.\vec{x}}\,a_{\vec{k}}^\dagger\right], 
\ee
where $\vec{k}$ is the comoving momentum variable, $a_{\vec{k}}$ and $a^\dagger_{\vec{k}}$ denote the  {\it time-independent} ladder operators obeying $[a_{\vec{k}},a^\dagger_{\vec{k}'}]=\d(\vec{k}-\vec{k}')$.  Furthermore, the mode functions satisfy  the Wronskian condition 
\be\label{r}
\m_k\m_k'^*-\m_k^*\m_k'=i,
\ee
together with 
\be
\m_k''+\left[k^2-\fr{a''}{a}\right]\m_k=0.\label{mf}
\ee
The ground state $|0>$  at time $\eta_0$ can be defined by imposing 
\be
a_{\vec{k}}\,|0>=0.
\ee
The quantum system is uniquely specified if one additionally  gives the initial values $\m_k(\eta_0)$ and  $\m_k'(\eta_0)$ obeying the Wronskian condition \eq{r}. 

Using the definition of the stress-energy-momentum tensor $T_{\m\n}=\nabla_\m\phi \nabla_\n\phi-\fr{1}{2}g_{\m\n}(\nabla\phi)^2$, one can  calculate the following vacuum-expectation values 
\bea
&&<0|\r|0>=\frac{1}{4\pi^2 a^4}\int_0^\infty k^2\left[ a^2
\left|\left(\fr{\m_k}{a}\right)'\right|^2+k^2|\m_k|^2\right]dk,\nn\\ 
&&<0|P|0>=\frac{1}{4\pi^2 a^4}\int_0^\infty k^2\left[ a^2
\left|\left(\fr{\m_k}{a}\right)'\right|^2-\fr{k^2}{3}|\m_k|^2\right]dk.
\label{se} 
\eea
These expressions contain quartic, quadratic and logarithmic divergences. Adiabatic regularization \cite{ad1} offers a physically viable way of removing these divergences using suitably defined subtraction terms.  To determine the adiabatic subtraction terms, one defines the adiabatic mode function $\m^{ad}_k$ in terms of a new variable $\O_k$  in the following WKB form  
\be
\m^{ad}_k=\fr{1}{\sqrt{2\O_k}}\,e^{-i\int \O_k d\eta}. \label{admf} 
\ee
Using \eq{mf}, $\O_k$ can be seen to obey  
\be\label{11}
\O_k^2=k^2-\fr{a''}{a}+\fr{3}{4}\fr{\O_k'^2}{\O_k^2}-\fr{1}{2}\fr{\O_k''}{\O_k}.
\ee
Eq. \eq{11} can be solved iteratively, where the number of time derivatives acting on $\O_k$ is used as an expansion parameter. The zeroth order adiabatic solution is $\O_k=k$ and using it in the right hand side of \eq{11} the second order adiabatic solution  can be found as $\O_k=k\left[1-\fr{a''}{2ak^2}\right]$. To regularize \eq{se}, i.e. to  remove  quartic, quadratic and logarithmic divergences, one should determine $\O_k$ up to adiabatic order four terms and subtract the corresponding stress-energy tensor expressions obtained from the adiabatic mode function \eq{admf}. After a straightforward calculation, the regularized finite expressions for the vacuum expectation values can be found as 
\bea
\r_V&=&\frac{1}{4\pi^2 a^4}\int_0^\infty k^2\left[ a^2
\left|\left(\fr{\m_k}{a}\right)'\right|^2+k^2|\m_k|^2-k-\fr{h^2}{2k}-\fr{1}{8k^3}(3h^4+h'^2-2hh'')\right]dk,
\label{frho}\\ 
P_V&=&\frac{1}{4\pi^2 a^4}\int_0^\infty k^2\left[ a^2
\left|\left(\fr{\m_k}{a}\right)'\right|^2-\fr{k^2}{3}|\m_k|^2-\fr{k}{3}-\fr{h^2}
{6k}+\fr{h'}{3k} \right.\nn\\
&&\hs{50}\left. -\fr{1}{24k^3}(3h^4+h'^2-2hh''-12h^2h'+2h''') \right]dk.\label{fp}
\eea
One may check that $\r_V$ and $P_V$ obey the conservation equation $\rho_V'+3h(\rho_V+P_V)=0$ provided \eq{mf} holds. 
 
There is an important subtlety in using adiabatic regularization for massless fields. If one naively performs adiabatic regularization for a {\it conformally coupled} massless scalar field, the regularized stress-energy tensor vanishes identically, i.e. it is not possible to obtain the  conformal anomaly. The correct procedure is to start with a massive field and then send the mass parameter to zero by carefully keeping the terms that survives the limit. Only in this way one can get the conformal anomaly using adiabatic regularization \cite{tr}. In our case, the same procedure should be used, this time for a minimally coupled field. These terms can be determined by using the formulas given in \cite{ad-f}.  Similarly, in \cite{ali2} we also find out  the adiabatic subtraction terms for a massive field up to sixth order. From our calculation in \cite{ad-f} we note the fourth order adiabatic subtraction terms for energy density as 
\bea
&&\int_0^\infty  \fr{k^2\,dk}{512\pi^2a^7\o^{11}}\left[56m^6\o^2a^6a'^2a''-105 m^8 a^7a'^4+144m^4\o^4a^4a'^2a''+16m^2\o^6a^3(a''^2-2a'a''')
\right.\nn\\
&&+64\o^8a'^2a''-224m^6\o^2a^5a'^4+4m^4\o^4a^5(a''^2-2a'a''')+64\o^{10}a a'^2+16m^2\o^6aa'^4 +128\o^{12}a^3\nn\\
&&\left.+16m^4\o^6a^5a'^2+16\o^8a(a''^2-2a'a''')+64m^2\o^8a^3a'^2+124m^4\o^4a^3a'^4
+96m^2\o^6a^2a'^2a''\right],\nn
\eea
where $\o=\sqrt{k^2+m^2 a^2}$. By scaling $k\to mk$ the integrals of the terms in the square brackets with $m^8$, $m^6\o^2$, $m^4\o^4$ and $m^2\o^6$  can be seen to be $m$-independent and finite (note that the integral of $\o^8$ term is also mass independent but it diverges). Performing the integrals one can find  the finite adiabatic subtraction terms for energy density
\be\label{rff}
\r_F=\fr{1}{960\pi^2a^4}\left[92h^4+60h^2h'+11h'^2-22hh''\right].
\ee
The corresponding expression for  pressure can be determined from the stress-energy conservation $\rho_F'+3h(\rho_F+P_F)=0$, which gives
\be\label{pff}
P_F=\fr{1}{2880\pi^2a^4}\left[92h^4-308h^2h'-109h'^2-82hh''+22h''' \right].
\ee
These are the extra terms which must be  subtracted from \eq{frho} and \eq{fp} 
\bea
&&\r_V \to \r_V-\r_F,\label{rf}\\
&&P_V\to P_V-P_F.\label{pf}
\eea

One can  use  $\r_V$ and $P_V$ as sources in the Einstein equations:
\bea
&& h^2=\fr{8\pi a^2}{3M_p^2}\,\r_V,\label{e1}\\
&& h'=-\fr{4\pi a^2}{3M_p^2}\,(\r_V+3P_V).\label{e2}
\eea
The evolution is  fully specified once the initial conditions for the mode function $\m_k$ and the scale factor $a(t)$ are given. This is an integro-differential equation system: the integrals of the mode functions determine the stress-energy tensor as in  \eq{frho} and \eq{fp}, and these are used in Einstein equations to determine the dynamics of the scale factor. We have also \eq{mf}, which fixes the evolution of the mode functions. 

Although \eq{frho} and \eq{fp} are guaranteed to be {\it ultraviolet} finite for suitably chosen mode function initial conditions, there are potentially dangerous infrared divergences for massless fields \cite{ir}.  For example, when $a=(\eta_0/\eta)^n$ the mode function for the Bunch-Davies vacuum is given by $\m_k=\eta\, \sqrt{k/2}\, h_n(k\eta)$, where $h_n$ is the spherical Hankel function of first kind. From the behavior of $h_n(k\eta)$ as $k\to 0$, one can see that the integrals of the mode functions in \eq{frho} and \eq{fp} diverge with a power near $k=0$. Similarly, the integrals of the fourth order adiabatic  subtraction terms in \eq{frho} and \eq{fp} (these are the terms with $1/k^3$  in the square brakets) diverge logarithmically near $k=0$. Therefore, one should introduce an infrared cutoff to make sense of $\r_V$ and $P_V$. This is similar to the  point splitting regularization, which also requires an infrared cutoff for massless fields. 

\section{An approximation}

In this section we would like to utilize an approximation to simplify the integrals in \eq{frho} and \eq{fp}. First, we assume that the field is placed in Bunch-Davies vacuum or $l$'th order adiabatic vacuum with $l\geq4$ such that as $k\to \infty$ one discovers usual mode functions of the Minkowski space.\footnote{One should note that in realistic situations there is a certain uncertainty in specifying the vacuum \cite{min}. See also \cite{kundu} for possible cosmological implications of  different initial states in inflation.} The differential equation determining the evolution of the mode functions \eq{mf} can be solved in two different limits: $\m_k\simeq e^{ ik\eta}/\sqrt{2k}$ if $k\gg h$ and $(\m_k/a)'\simeq 0$ if  $k\ll h$, respectively, corresponding to the subhorizon and the superhorizon regimes.

To simplify \eq{frho} and \eq{fp} we use the fact that subhorizon modes evolve adiabatically and superhorizon modes freeze-out. Therefore, in the chosen vacuum the difference between the actual mode function $\m_k$ and the adiabatic mode function $\m^{ad}_k$ in \eq{admf} should be negligible for subhorizon modes. Aditionally, we take superhorizon modes to obey exactly $(\m_k/a)'= 0$. To have a better approximation, we introduce a real parameter $\a$, which quantifies subhorizon/superhorizon border. Namely for $k>\a h$ we ignore the difference between  $\m_k$ and $\m^{ad}_k$ in the integrals in \eq{frho} and \eq{fp}, and for $k<\a h$ we take  $(\m_k/a)'= 0$.  The need for such a parameter can be justified as follows. From the differential equation \eq{mf}, subhorizon/superhorizon regimes are determined by $\sqrt{a''/a}$, which may be different than $h$. Moreover, there appears errors in the integrals for using $\m^{ad}_k$ instead of $\m_k$ for subhorizon modes and imposing  $(\m_k/a)'= 0$ exactly for superhorizon modes. The $\a$-parameter is introduced to compensate these errors and it is expected to be of order unity.\footnote{Our $\a$ parameter is similar to the $\e$ parameter  of Starobinsky introduced in \cite{str}.}

One can  evaluate \eq{frho} and \eq{fp} approximately as follows. First, the integrals are negligible in the $[\a h,\infty]$ range by the above argument. To calculate the integrals in $[0,\a h]$ interval we define the following variables 
\bea
&&\r_S=\frac{1}{4\pi^2 a^4}\int_{0}^{\a h} k^2\left[ a^2
\left|\left(\fr{\m_k}{a}\right)'\right|^2+k^2|\m_k|^2\right]dk,
\label{rs}\\ 
&&P_S=\frac{1}{4\pi^2 a^4}\int_{0}^{\a h} k^2\left[ a^2
\left|\left(\fr{\m_k}{a}\right)'\right|^2-\fr{k^2}{3}|\m_k|^2 \right]dk.\label{rp}
\eea
Since we approximate superhorizon modes to obey\footnote{In the superhorizon regime,  there are  two solutions to \eq{mf}: $\m_k(\eta)\simeq c_1(k)a(\eta)+c_2(k)a(\eta)\int_{\eta_0}^{\eta}d\eta'/a^2(\eta')$, where \eq{r}  implies $c_1c_2^*-c_1^*c_2=i$. The second solution is the decaying mode and it becomes negligible in time. Indeed, $(\m_k/a)'\simeq c_2/a^2$ and thus first terms in \eq{rs} and \eq{rp} decrease $1/a^2$ more, which justifies the equation of state \eq{eos}.}  $(\m_k/a)'=0$, these variables satisfy
\be \label{eos}
P_S=-\fr13 \rho_S.
\ee
Performing the integrals \eq{frho} and \eq{fp} one  finds  
\bea
\r_V&=&\r_S-\fr{1}{4\pi^2 a^4}\left[\fr{(\a^4+\a^2)}{4} h^4+\fr18 \ln\left(\fr{\a h}{h_0}\right)(3h^4+h'^2-2hh'')\right] -\r_F,\label{v1}\\
P_V&=&P_S-\fr{1}{4\pi^2 a^4}\left[\fr{(\a^4+\a^2)}{12} h^4-\fr{\a^2}{6} h^2h'\right. \nn\\
&&\left.+\fr{1}{24}\ln\left(\fr{\a h}{h_0}\right)(3h^4+h'^2-2hh''-12h^2h'+2h''')\right]-P_F,
\eea
where a comoving infrared cutoff $h_0$ is introduced to make sense of the logarithmic integrals (the cutoff must be comoving in order to preserve stress-energy conservation). Infrared divergences, which might be related to the behavior of the mode functions, are absorbed in the definitions of $\r_S$ and $P_S$. Thus, the vacuum expectation value of the stress-energy tensor is determined up to an unknown function $\r_S$, which is actually fixed by the initially chosen vacuum state.   

The combination $\r_V+3P_V$, which appears in the right hand side of the acceleration equation \eq{e2}, becomes
\bea
\r_V+3P_V&=&-\fr{1}{4\pi^2 a^4}\left[\fr{(\a^4+\a^2)}{2} h^4-\fr{\a^2}{2} h^2h'\right. \nn\\
&&\left.+\fr{1}{8}\ln\left(\fr{\a h}{h_0}\right)(6h^4+2h'^2-4hh''-12h^2h'+2h''')\right]-(\r_F+3P_F). \label{r3p}
\eea
Thus, $\r_S$ dependence disappears and \eq{e2} can be studied  to determine the  evolution of the scale factor $a(\eta)$. 

Note that $\r_V$ and $P_V$ must obey stress-energy conservation. This is guaranteed for the exact expressions \eq{frho} and \eq{fp}, and we can now impose it to determine the evolution of the vacuum energy density. Plugging \eq{v1} to the right hand side of $\rho_V'=-3h(\rho_V+P_V)$, using $P_S=-\rho_S/3$ and further solving for $\r_S$ in terms of $\r_V$ from \eq{v1} one finds\footnote{Recall that $\r_F$ and $P_F$ are given in \eq{rff} and \eq{pff}. These also obey stress-energy conservation, which we use in \eq{son} to substitute for $\r_F'$.}
\bea
\r_V' + 2h \r_V&=&\fr{1}{4\pi^2 a^4}\left[\fr{(\a^4+\a^2)}{2} h^5-\fr{\a^2}{2}h^3h'\right.\nn\\
&+&\left. \fr18 \ln\left(\fr{\a h}{h_0}\right)(6h^5+2hh'^2-4h^2h''+2hh'''-12h^3h')\right]
+h (\r_F+3P_F). \label{son}
\eea
This is our final expression for the vacuum energy density. The backreaction problem is  expressed in terms of two variables, namely the scale factor of the universe $a(\eta)$ and the vacuum energy density $\r_V$,  which obey  the Friedmann equation \eq{e1} and the conservation equation \eq{son}. As usual the acceleration equation \eq{e2} is satisfied identically once \eq{e1} and \eq{son} are hold. 

Since $\r_V$ obeys a first order differential equation, its initial value must be  specified, which is fixed by the initial vacuum chosen. Introducing the homogeneous and particular solutions $\r_V=\r_H+\r_P$, \eq{son} implies that the homogeneous piece evolves as $\r_H=C_0/a^2$, where the constant $C_0$ can be related to the initial value of $\r_V$. The decrease of the homogeneous piece is equivalent to a perfect fluid with an equation of state parameter $\o=-1/3$, which is slower than radiation and dust. 

In general, $\a$ parameter is expected to depend on time but its time dependence must be  weak since $\a$ is of order unity. In any case, it is not possible to integrate \eq{son} to fix $\r_V$ exactly. However, one can estimate the magnitude of the vacuum energy density and determine how it evolves. 

The comoving infrared cutoff $h_0$ must also be chosen appropriately. One way of dealing with the infrared problem is to assume that the vacuum has no superhorizon modes initially (see e.g. \cite{irw}). Thus $h_0$ can be identified as the initial Hubble radius in conformal time.

It is important to note that for the above approximation to work there must not exist  flow of modes from superhorizon to subhorizon regime, since for these modes the difference between 
the actual mode function $\m_k$ and  the adiabatic mode function $\m_k^{ad}$ will not be small (recall that superhorizon modes evolve non-adiabatically). Therefore, the above formulas can safely be used only for accelerating cosmologies. However, by comparing \eq{son} with the known results,  we will show below that there is good agreement even for decelerating cosmologies. In these cases the $\a$ parameter compensates the errors and the method works fine even for decelerating cosmologies.

To check the validity of our method assume that the background evolution is fixed, i.e. the scale factor $a(\eta)$ is given. Then, \eq{son} dictates how vacuum energy density evolves in time, which can be compared with the known results in the literature. 

Let us start with de-Sitter space and take
\be 
a=\fr{\eta_0}{\eta}. 
\ee
The vacuum energy density of a minimally coupled massless scalar field in the Bunch-Davies vacuum was calculated in \cite{r-ds}, which can also be found in the book \cite{b1}:
\be \label{dsr}
\r_{BD}=\fr{-29h^4}{960\pi^2a^4}.
\ee
Integrating \eq{son} we find
\be 
\r_V=\fr{(-119+60\a^4)h^4}{960\pi^2a^4}+\fr{C_0}{a^2}.
\ee
Therefore, for $C_0=0$ and $\a=(3/2)^{1/4}$ our calculation exactly matches the well-known result in de Sitter space. Since Bunch-Davies vacuum is de Sitter invariant the vacuum energy density must be a constant, which would justify the choice $C_0=0$. On the other hand, $\a$ parameter turns out to be of order one as expected.   

\begin{table}
\centering
\begin{tabular} {|c | c|}
\hline
\hs{3}$-\infty <n<0$\hs{3} & \hs{3}Decelerating\hs{3} \\
\hline
$n=0$ & Minkowski \\
\hline
$0<n<1$ & Big-crunch \\
\hline
$n=1$ & de Sitter \\
\hline
$1<n$ & Accelerating \\
\hline
\end{tabular}
\caption{Classification of the universes with power law expansion in the conformal time. The scale factor \eq{rw} in the proper time is given by $a=(t/t_0)^m$, where $m=n/(n-1)$.}
\end{table} 

Consider now a background with power-law expansion
\be\label{rw}
a=\left(\fr{\eta_0}{\eta}\right)^n. 
\ee
Using point-splitting regularization, the stress-energy tensor of a minimally coupled massless quantum scalar field in the Bunch-Davies vacuum has been calculated in \cite{bdrw}, where the vacuum energy density is given by 
\bea
\r_{BD}&=&\fr{1}{2880\pi^2}\left[-\fr16   {}^{(1)}H_0{}^0+{}^{(3)}H_0{}^0\right]-\fr{1}{1152\pi^2}{}^{(1)}H_0{}^0\left[\ln\left(\fr{R}{\m^2}\right)+\psi(2+n)+\psi(1-n)+\fr43\right]\nn\\
&&+\fr{1}{13824\pi^2}\left[-24\Box R+24R R_0{}^0+3R^2\right]-\fr{R}{96\pi^2a^2\eta^2}.\label{31}
\eea
Here $\psi$ is the di-gamma function, $\m$ is an arbitrary constant  and 
\bea
&&{}^{(1)}H_{\m\n}=2\nabla_\m\nabla_\n R-2g_{\m\n}\Box R-\fr12 g_{\m\n}R^2+2RR_{\m\n},\\
&&{}^{(3)}H_{\m\n}=R_{\m}{}^{\r}R_{\r\n}-\fr32 RR_{\m\n}-\fr12 g_{\m\n}R_{\r\s}R^{\r\s}+\fr14 R^2g_{\m\n}.\label{h3}
\eea
Scaling $\m$ is equivalent to finite renormalization of a coupling constant in the quantum effective action (see e.g. comments below eq. (7.52) in \cite{b1}). Therefore, one may actually add an arbitrary multiple of ${}^{(1)}H_0{}^0$ to $\r_{BD}$. In this way, the extra divergences that may arise when the argument of the di-gamma function is zero or negative integer can also be removed. 

Using \eq{rw}, we evaluate \eq{31} as 
\bea 
\r_{BD} &=&\fr{h^4}{1920\pi^2n^2a^4}\left[81-30n-109n^2\right.\nn\\
&-&\left. 90(n^2-1)\left(\ln\left[\fr{6(n+1)h^2}{n\m^2}\right]+\psi(1-n)+\psi(n+2)\right)\right], \hs{5}n\not=-1.\label{rbd}
\eea
When $n=-1$, which corresponds to a radiation dominated universe, ${}^{(1)}H_0{}^0$ vanishes identically, so  there does not arise any issue with the logarithm and the di-gamma functions. In that case, the vacuum energy density can be found as
\be\label{a1}
\r_{BD}=\fr{h^4}{960\pi^2a^4},\hs{10}n=-1.
\ee
To avoid the subtleties related to the singularities of the di-gamma function,  it is safe to take $n$ to be a {\it non-integer}. 

The radiation dominated background $n=-1$ is special since the exact mode function can be determined from \eq{mf} as $\m_k=e^{-ik\eta}/\sqrt{2k}$, where the proper normalization is chosen  for the Bunch-Davies vacuum. One then finds that \eq{frho} and \eq{fp} vanish identically. On the other hand, the finite adiabatic subtraction terms \eq{rff}  can de determined as  $\r_F=-h^4/(960\pi^2a^4)$, which gives
\be\label{a2}
\r_{V}=\fr{h^4}{960\pi^2a^4}.
\ee
The agreement between \eq{a1} and \eq{a2} shows the necessity of keeping finite adiabatic subtraction terms for massless fields. In terms of our approximation for evaluating the integrals \eq{frho} and \eq{fp}, eq. \eq{a2} corresponds  $\a=0$, which may be considered as a singular case. 

Integrating \eq{son} for the background \eq{rw} we find
\bea
\r_V&=&\fr{h^4}{1920\pi^2n^2a^4}\left[\fr{2}{n-2}\left(-21+153n+(79+60\a^2)n^2-(92+60\a^2+60\a^4)n^3\right)\right.\nn\\
&-& \left. 90(n^2-1)\ln\left[\fr{\a^2 h^2}{h_0^2}\right]\right]+\fr{C_0}{a^2}.\label{rv}
\eea
By comparing \eq{rv}  with \eq{rbd}, the Bunch-Davies vacuum can be seen to correspond $C_0=0$. The coefficient of $\ln(h^2)$  in \eq{rv} is $\a$-independent and  this term agrees exactly with \eq{rbd}, which is highly non-trivial since one cannot adjust any free parameters.  The other terms can actually be made to agree for any value of $\a$ by choosing the ratio $\m/h_0$   appropriately (as noted above this might be interpreted as a finite renormalization of a coupling constant in the effective action). 

\begin{figure}
\centerline{
\includegraphics[width=5cm]{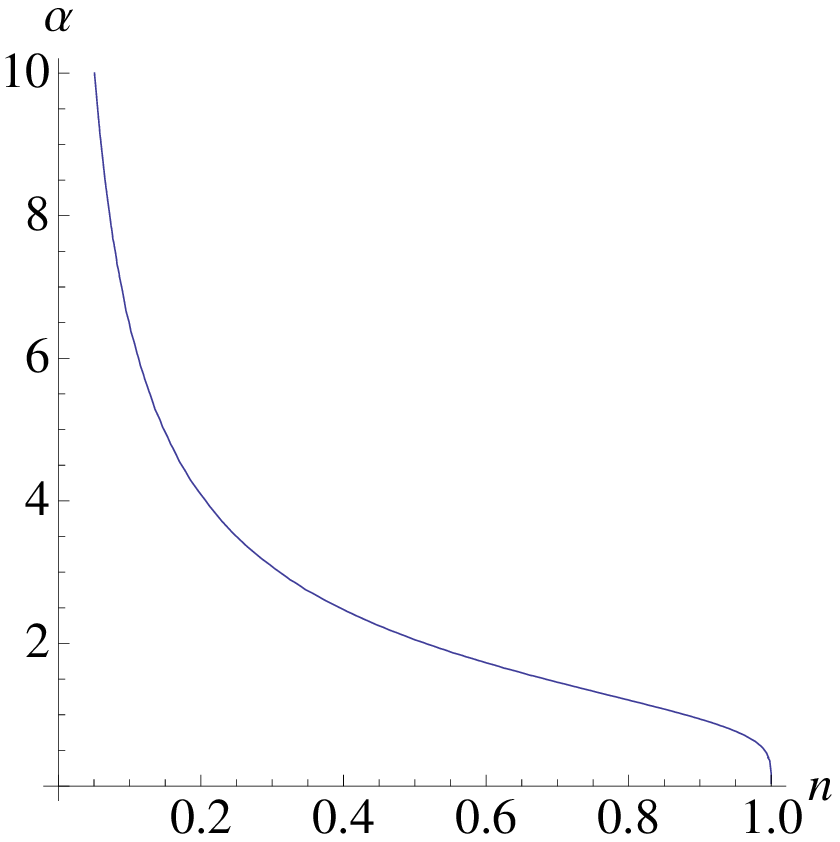} \includegraphics[width=5cm]{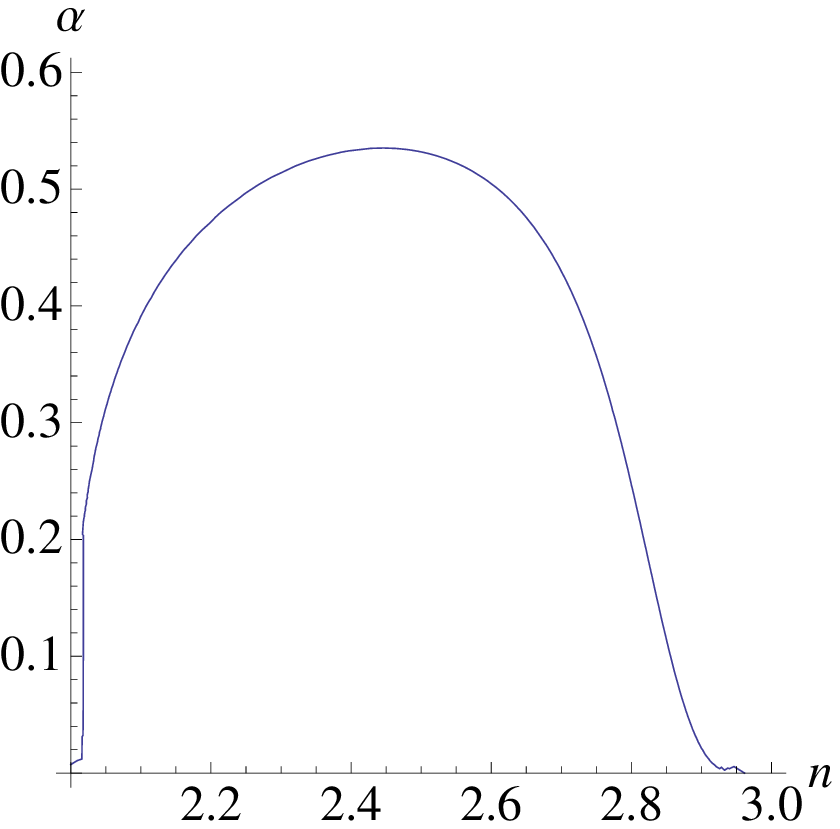} \includegraphics[width=5cm]{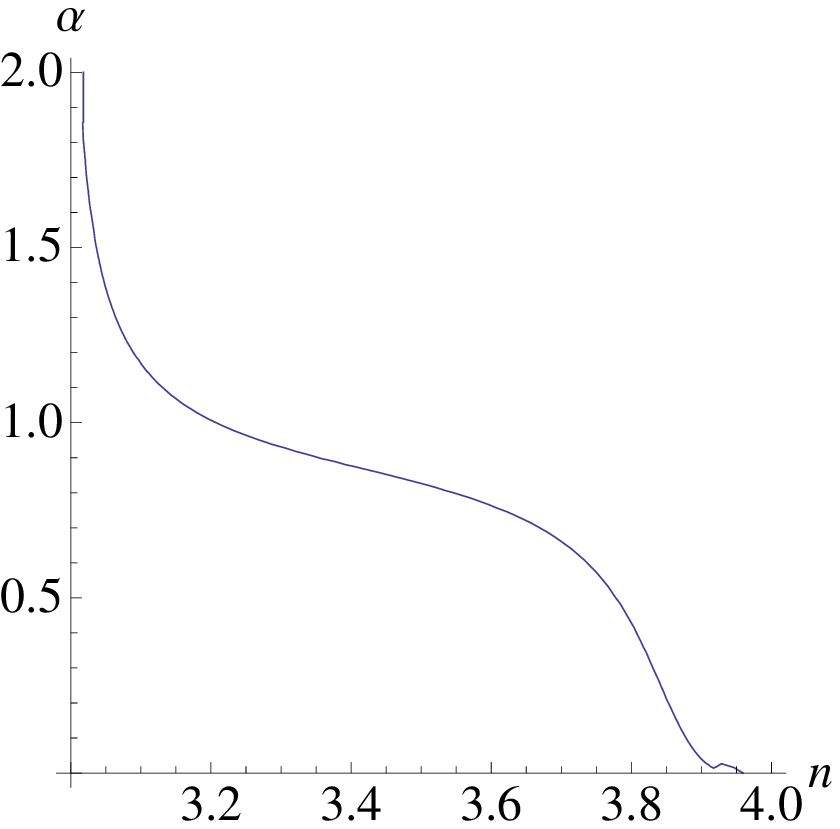} }
\centerline{\includegraphics[width=5cm]{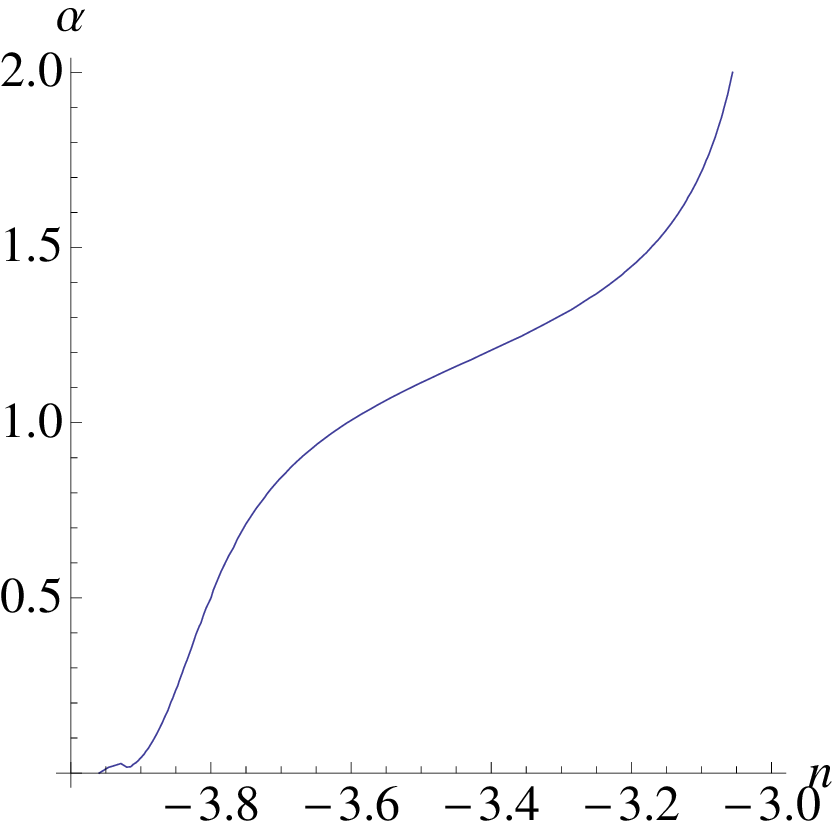} \includegraphics[width=5cm]{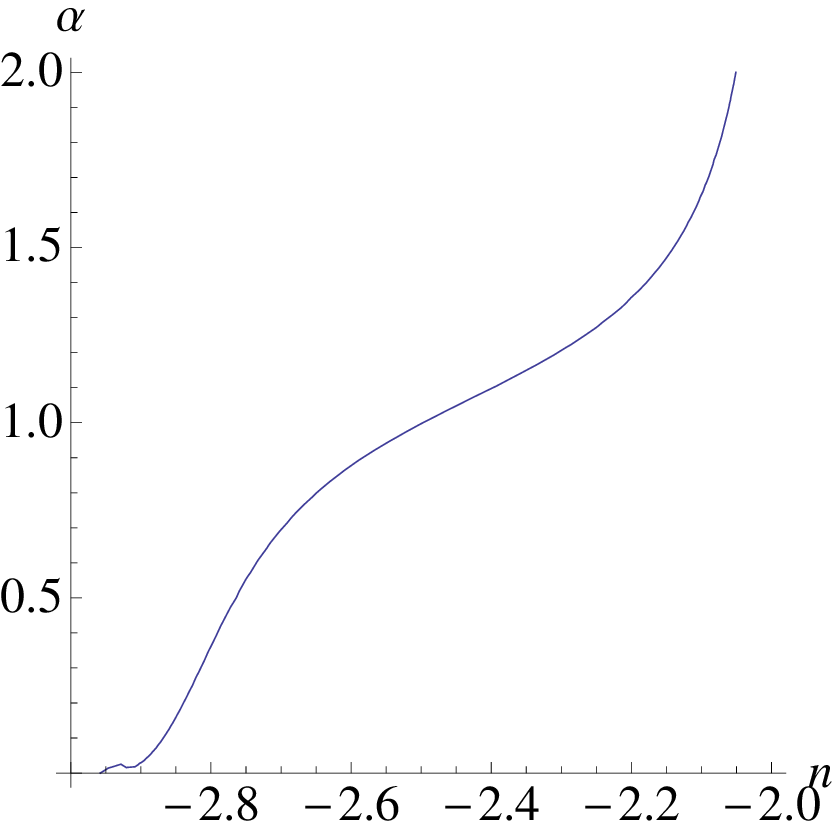} \includegraphics[width=5cm]{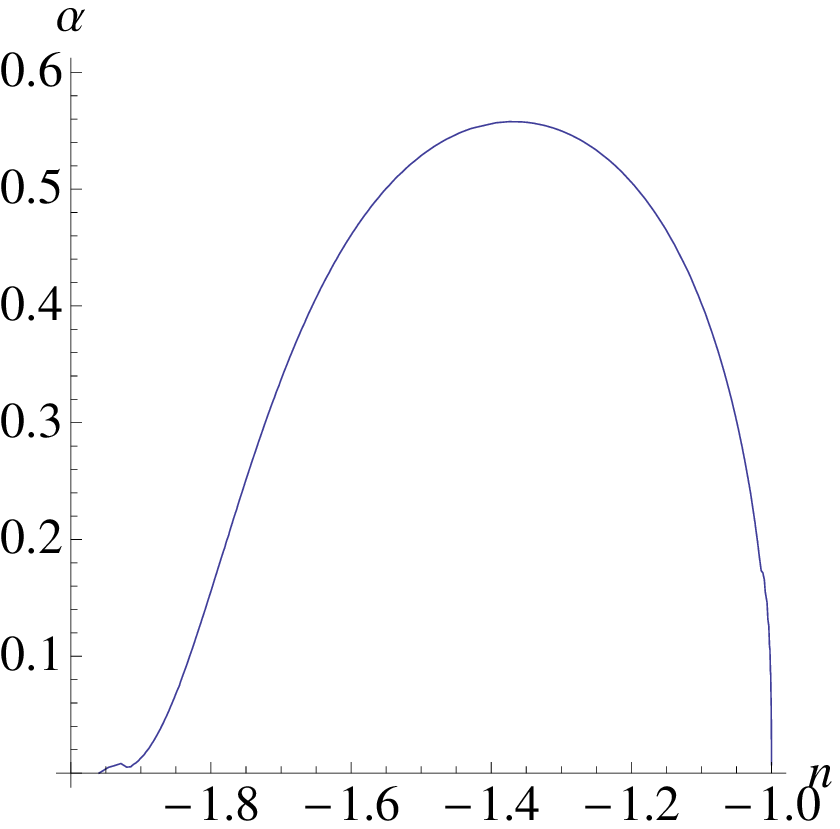}}
\caption{The numerical plots of $\a$-parameter as a function of the power $n$ from \eq{an}. For larger $|n|$ the plots look very similar to $(3,4)$ or $(-4,-3)$ intervals for positive and negative values, respectively. } 
\label{fig1}
\end{figure}

One may insist  to set 
\be
\m=h_0,
\ee
 since both of these parameters are  infrared cutoffs. In that case, to be able to get $\r_V=\r_{BD}$ the following equation must be satisfied: 
\be\label{an}
\ln\left[\fr{36(n+1)^2}{n^2\a^4}\right]+2\psi(1-n)+2\psi(n+2)
 =\fr{(5+8\a^2+8\a^4)n^3+(2-8\a^2)n^2-11n-8}{3(n^2-1)(n-2)}. 
\ee
It is not possible to solve $\a$ from this equation analytically.  In Fig. \ref{fig1} we give numerical plots for different intervals of unit size (as noted below \eq{h3}, for integer values of $n$ an infinite renormalization is required to make sense of $\r_{BD}$). As $n\to \pm\infty$,  \eq{an} reduces to 
\be
6\ln(\a^2/6)+(5+8\a^2+8\a^4)=2\psi(1-n)+2\psi(n+2).
\ee
Using $\psi(x)-\psi(x-1)=1/(x-1)$, one sees that the solution for $\a$ in the interval $(n,n+1)$ becomes independent of $n$ for large $|n|$. From the graphs it is obvious that $\a$ remains to be of order unity for generic values, which shows the validity of our approximation. On the other hand, for $n\in (1,2)$ and $n\in (-1,0)$ there turns out to be no solution for $\a$. This last fact together with the observation that as $n$ approaches to an integer $\a$ increases and in the limit ceases to be of order unity signal the need for a finite renormalization such that $\m\not = h_0$. 

\begin{figure}
\centerline{
\includegraphics[width=6cm]{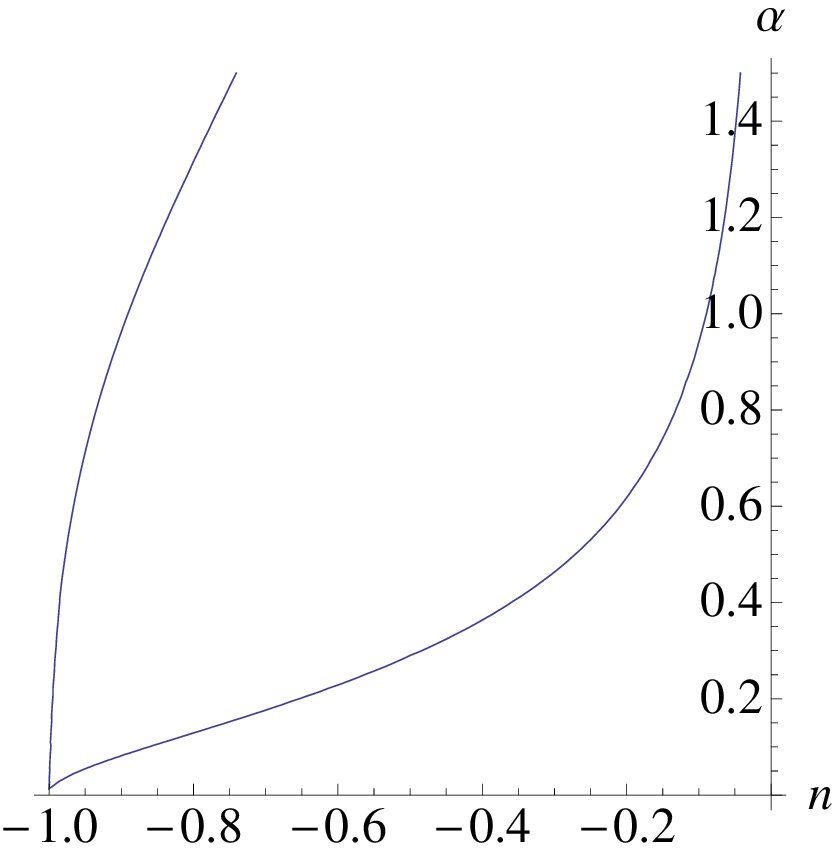}\hs{10} \includegraphics[width=6cm]{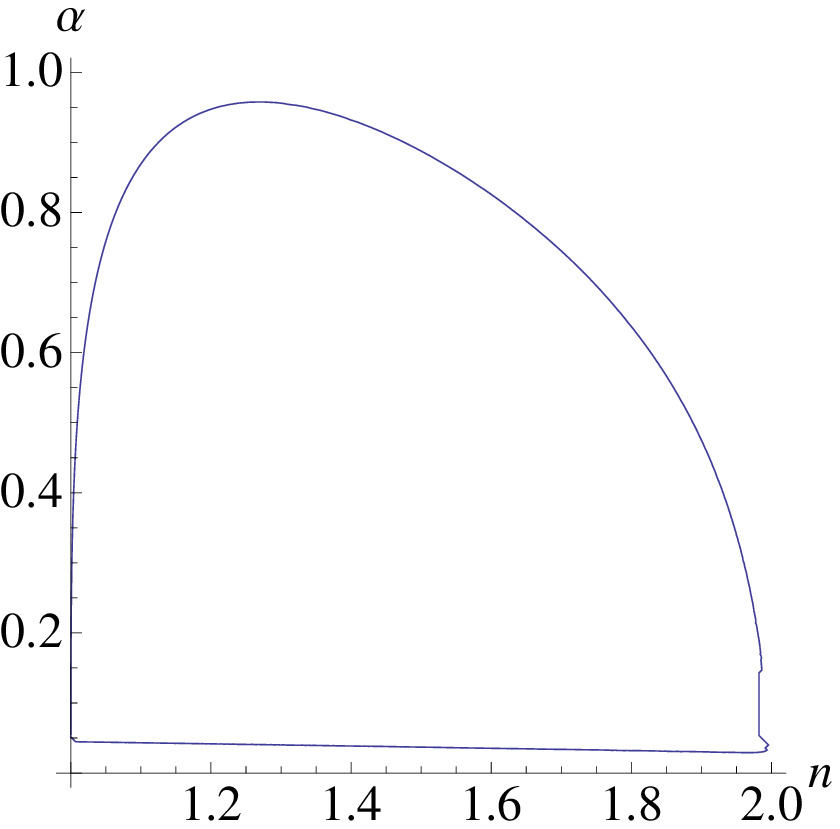} }
\caption{The numerical plots of $\a$-parameter as a function of the power $n$ for the interval $(-1,0)$, where $\m=10h_0$ and for the interval $(1,2)$, where $\m=100h_0$.} 
\label{fig2}
\end{figure}

When $n\in (-1,0)$, if one chooses $\m=10 h_0$, which corresponds to $\a\to 10\a$ in \eq{son}, there exists solutions for $\a$. Similarly, for $n\in (-1,0)$, if one chooses $\m=100 h_0$, which corresponds to $\a\to 100\a$ in \eq{son}, $\a$ becomes soluble. Corresponding numerical plots are given in Fig. \ref{fig2}. Note that for any given $n$ in these intervals there appears two solutions of $\a$. 

In summary,  we see that using our approximation it is possible to reproduce the known results in the literature. In de Sitter space ($n=1$), it is enough to choose $\a=(3/2)^{1/4}$. The  radiation dominated universe ($n=-1$) is singular since the result is given solely by the finite adiabatic subtraction terms, which corresponds to the choice $\a=0$. These are the only two cases with ${}^{(1)}H_0{}^0=0$. For all other integer values of $n$, one needs to apply an infinite renormalization to \eq{rbd} by scaling the constant $\m$. We see that the logarithmic term, which is  not affected by finite renormalizations, matches exactly in \eq{rbd} and \eq{rv}. The others with non-integer $n$  can make to agree either by scaling $\m$ or by adjusting $\a$. It turned out that the approximation also works for negative values of $n$, i.e. for decelerating cosmologies, see Fig. \ref{fig1}. As discussed above, for decelerating backgrounds extra care is needed for the modes, which was born as superhorizon and becomes subhorizon in time. The error appeared for neglecting these modes is compensated by the $\a$-parameter. 

\section{Application to slow-roll inflation}

In this section we would like to apply our findings to the simplest inflationary scenario realized by a single inflaton field $\vf$ with a suitable potential $V(\vf)$. Our aim is to see whether the backreaction of a free massless quantum field $\phi$ has any impact on the standard inflationary results. First, let us review the well-known results by ignoring the backreaction effects. As usual, the stress-energy tensor of the inflaton field is given by 
\bea
&&\r_\vf=\fr12 \dot{\vf}^2+V(\vf),\\
&&P_\vf=\fr12 \dot{\vf}^2-V(\vf).
\eea
Recall that the dot denotes differentiation with respect to the proper time $t$ defined as $dt=ad\eta$. The inflaton obeys the scalar equation 
\be\label{ie}
\ddot{\vf}+3H\dot{\vf}+\del_\vf V=0. 
\ee
To realize inflation in the slow-roll regime, one must assume that 
\be
\dot{\vf}^2\ll V,\hs{10}  \ddot{\vf}\ll 3H\dot{\vf}.
\ee
To satisfy these conditions it is enough to take the slow-roll parameters $\e_\vf$ and $\eta_\vf$ to be small: 
\bea
&&\e_\vf=\fr{M_p^2}{16\pi}\left(\fr{\del_\vf V}{V}\right)^2\ll1,\\
&&\eta_\vf=\fr{M_p^2}{8\pi}\left(\fr{\del^2_\vf V}{V}\right)\ll1. 
\eea
In this regime the inflaton equation becomes
\be
\dot{\vf}\simeq -\fr{\del_\vf V}{3H}.
\ee
Moreover, the Friedmann equation implies
\be
H^2\simeq \fr{8\pi}{3M_p^2}V.
\ee
If one defines the slow-roll parameters characterizing the change in the Hubble parameter and the inflaton as
\be\label{e}
\dot{H}\simeq-\e H^2,\hs{10} \ddot{\vf}\simeq -(\eta-\e)H\dot{\vf},
\ee 
it is possible to show that
\be
\e=\e_\vf,\hs{10}\eta=\eta_\vf.
\ee
Note that $\e$ and $\eta$ determine the tilt in the spectrum of the curvature perturbation and they are observationally important parameters of inflation. 

Consider now the backreaction of the quantum field $\phi$. From the differential equation \eq{son} satisfied by $\r_V$, one can estimate that
\be
\r_V={\cal O}(H^4). 
\ee 
Indeed, in the exact de Sitter space the vacuum energy density in the Bunch-Davies vacuum is given by \eq{dsr}. In the slow-roll regime one would not expect a large deviation from \eq{dsr}. In the following, we will treat $\e_\vf$, $\eta_\vf$ and 
\be
\cc\equiv \fr{H^2}{M_p^2}
\ee
as small parameters and keep only the leading order terms. Since we do not assume any coupling between $\phi$ and $\vf$, the inflaton equation does not change:
 \be\label{sr} 
\dot{\vf}\simeq -\fr{\del_\vf V}{3H}.
\ee
On the other hand, from the Friedmann equation
\be\label{fr}
H^2=\fr{8\pi}{3M_p^2}(\r_V+\r_\vf),
\ee
one sees that $\r_\vf\sim H^2M_p^2$ and thus $\r_V\sim \cc \r_\vf$.  Using  \eq{r3p}, \eq{rff} and \eq{pff} in Einstein's equations, keeping only the leading order terms in the slow-roll parameters 
 we  get 
\be\label{1}
\dot{H}=-\fr{8\pi}{3M_p^2}\r_V-\left[\fr{119}{360\pi}-\fr{\a^4}{6\pi}\right]\fr{H^4}{M_p^2}-\fr{4\pi}{M_p^2}\dot{\vf}^2.
\ee
From \eq{sr} and the Friedmann equation \eq{fr},  we find  
\be
\dot{\vf}^2=\fr{M_p^2}{4\pi}\e_\vf H^2,
\ee
which is also true in the absence of the quantum scalar field (i.e. the presence of $\r_V$ in \eq{fr} does not alter the relation between inflaton kinetic energy and $H^2$ since $\e_\vf \cc$ is assumed to be negligible). Using this last equation in \eq{1} leads
\be\label{ne}
\e=\e_\vf +\left[\fr{119}{360\pi}-\fr{\a^4}{6\pi}+\fr{8\pi\r_V}{3H^4}\right]\cc,
\ee
where $\e$ is defined in \eq{e}. Similarly, taking the time derivative of \eq{sr} and using the definition of $\eta$ in \eq{e}, we find that
\be
\eta=\eta_\vf.
\ee
Therefore the backreaction of a quantum scalar field modifies the slow-roll parameter $\e$, i.e. it is not simply determined by the inflaton potential. 

It is easy to see that the change in $\e$  is actually very small. For example, considering a chaotic inflationary scenario with the potential $V=\fr12 m^2 \vf^2$, it is known that the cosmologically relevant perturbations exit the horizon when $\vf\sim 3 M_p-4 M_p$ and  a phenomenologically viable value of the inflaton mass is given by $m\sim 10^{-6}M_p$. One may then find that $\e_\vf\sim 10^{-3}$ and $\cc\sim 10^{-11}$, thus the modification due to backreaction effects is tiny. 

While the backreaction effects are negligible during the near exponential expansion, they might be important for the subsequent cosmic evolution. As noted above, \eq{son} shows  that the vacuum energy density, which  is driven by the terms in the right hand side, is of the order of $H^4$. However, the homogeneous piece decreases like $1/a^2$, which is  slower than  $H^4$ in radiation or dust dominated universes. Therefore, one expects $\r_V$ to become $H_0^4$ during the exponential expansion, where $H_0$ is the Hubble parameter of inflation, and then to decrease like $1/a^2$; thus one has $\r_V\sim H_0^4/a^2$ following inflation (this is indeed what is observed analytically in \cite{ali} in the simplified setting). On the other hand, the initial energy density of the radiation just after inflation can be estimated as $H_0^2M_p^2$, so the radiation energy density decreases like $\r_R\sim H_0^2M_p^2/a^4$. Comparing these two, we see that  when the universe expands $M_p/H_0$ times after inflation, the energy density of the vacuum catches up the energy density of radiation and then it starts dominating the universe. In the above mentioned chaotic inflationary model $M_p/H_0\sim 10^{6}$ and the equivalence corresponds to a very early cosmic epoch, for example, well before nucleosynthesis. The only loophole in the above conclusion is that \eq{son} is rigorously valid in an accelerating background. However, as we saw in the previous section \eq{son} gives correct results even for decelerating backgrounds by adjusting the $\a$-parameter. Therefore, our findings indicate that the vacuum energy density created during inflation must be taken into account for succeeding cosmic evolution. 

Of course, in a realistic model the evolution of the vacuum energy density must be studied by considering different stages carefully. For example, during the reheating or preheating stage the coupling of our scalar $\phi$ to the oscillating inflaton field may change the evolution of vacuum energy density substantially. Thus, it might be possible to recover the current standard cosmological picture by such modifications.  

\section{Conclusions}

Determining how  quantum fields affect the cosmological evolution, in the semiclassical approximation where gravity is treated classically, is an important problem. Quantum field theoretical effects can be crucial in resolving the initial big-bang singularity, during inflation or in explaining the presently observed acceleration of the universe. Compared to quantization in a fixed background, the backreaction problem is much more difficult to study since the mode functions of the field cannot be solved and thus their contributions to the stress-energy tensor cannot be determined explicitly. We manage to evaluate these contributions approximately by considering subhorizon and superhorizon modes separately. For a massless field, subhorizon modes with wavelengths smaller than the Hubble radius evolve adiabatically and their contributions cancel out  by the adiabatic subtraction terms. On the other hand, the contributions of the superhorizon modes to the stress-energy tensor obey an equation of state. In that way, we are able to obtain an explicit equation for the vacuum energy density. To have a better approximation, we introduce a parameter of order unity to quantify subhorizon/superhorizon division. By adjusting that parameter we  show that our approximation applied to a field propagating in a fixed background exactly agrees with the known results  in the literature. In our approach there are important subtleties related to field being massless. One should be careful about the  finite adiabatic subtraction terms which survive the massless limit. Moreover, there also exists infrared divergences which must be treated suitably. 

We apply our findings to a simple slow-roll inflationary scenario realized by a single scalar field. We observe that quantum fluctuations does not change the  background evolution appreciably during the near exponential expansion. They only induce a small change in one of the slow-roll parameters. However, vacuum energy density accumulated during the exponential expansion can be important at later stages, since it decreases like $1/a^2$, which is slower than radiation or dust. This decrease corresponds to an equation of state parameter $\o=-1/3$, therefore vacuum energy-density and vacuum pressure do not alter the acceleration equation for the scale factor of the universe following inflation.  

Encouraged by the fact that we can reproduce the known results in the literature by our approximation, the present work can be developed in different ways. It would be interesting to understand the situation with decelerating cosmologies better. For that, the contributions of the modes, which were born as superhorizon and later on become subhorizon, must be estimated. In that way, one can sharpen the conclusion that the vacuum energy density created during inflation becomes important at later stages of cosmic evolution. It is also worth to the consider the cosmological backreaction of a massive field. In that case, there exists two scales in the problem, the mass of the field and the Hubble parameter, and subhorizon/superhorizon separation of the modes is more complicated than that of the massless field (for instance if the mass of the field is larger than the Hubble radius, all modes evolve adiabatically). Moreover, for the contributions of the superhorizon modes to the stress-energy tensor, it seems difficult to write down an equation of state. Thus for a massive field one may attempt to use numerical analysis to solve the integro-differential equation system.

\end{document}